\documentclass[%
 reprint,
superscriptaddress,
showpacs,
 amsmath,amssymb,
 aps,
]{revtex4-1}

\usepackage{graphicx}
\usepackage{dcolumn}
\usepackage{bm}

\begin{document}

\preprint{CuFeO2}

\title{Incommensurate Orbital Modulation in Multiferroic CuFeO$_2$}

\author{Yoshikazu Tanaka}
\affiliation{%
 RIKEN SPring-8 Center, Sayo, Hyogo 679-5148, Japan}%
\author{Noriki Terada}%
\affiliation{%
National Institute for Materials Science, Sengen 1-2-1, Tsukuba, Ibaraki 305-0044, Japan
}%
\author{Taro Nakajima}%
\affiliation{%
Department of Physics, Faculty of Science, Tokyo University of Science, Tokyo 162-8601, Japan}%

\author{Taro Kojima}
\affiliation{%
 RIKEN SPring-8 Center, Sayo, Hyogo 679-5148, Japan}%
\affiliation{
Department of Complex Science and Engineering, The University of Tokyo, Kashiwa, Chiba 277-8561, Japan
}

\author{Yasutaka Takata}
\affiliation{%
 RIKEN SPring-8 Center, Sayo, Hyogo 679-5148, Japan}%
\affiliation{
Department of Complex Science and Engineering, The University of Tokyo, Kashiwa, Chiba 277-8561, Japan
}

\author{Setsuo Mitsuda}%
\affiliation{%
Department of Physics, Faculty of Science, Tokyo University of Science, Tokyo 162-8601, Japan}%

\author{Masaki Oura}                                                                                        
\affiliation{%
 RIKEN SPring-8 Center, Sayo, Hyogo 679-5148, Japan}%

\author{Yasunori Senba}
\author{Haruhiko Ohashi}
\affiliation{
Japan Synchrotron Radiation Research Institute (JASRI), Sayo, Hyogo
679-5198, Japan
}

\author{Shik Shin}
\affiliation{%
 RIKEN SPring-8 Center, Sayo, Hyogo 679-5148, Japan}%


\date{\today}

\begin{abstract}
CuFeO$_2$ is one of the multiferroic materials and is the first case that the electric polarization is not explained 
by the magnetostriction model or the spin-current model. 
We have studied this material using soft x-ray resonant diffraction and found that superlattice reflection 
0, $1-2q$, 0 appears
 in the ferroelectric and  incommensurate magnetic ordered phase at the Fe $L_{2,3}$ absorption edges
 and moreover that the rotation of the x-ray polarization such as 
 from $\sigma$ to $\pi$ or from $\pi$ to $\sigma$ is allowed at this reflection.
These findings definitely provide direct evidence that the $3d$ orbital state of Fe ions has 
a long range order in the ferroelectric state and support
 the spin-dependent \textit{d}-\textit{p} hybridization mechanism.  
\end{abstract}

\maketitle


The discovery of ferroelectricity induced by a magnetic order in TbMnO$_3$~\cite{Kimura2003mco} has attracted
 much attention for magnetoelectric (ME) multiferroic materials.
This new class of ferroelectric materials are essentially different from 
conventional ones like BiFeO$_3$~\cite{Wang2003ebm} where magnetism
 and ferroelectricity are independent to each other in principle.
The ferroelectricity found in the ME multiferroic materials appears as a result of the phase transition 
inducing such a magnetic order that breaks the crystal symmetry.
Geometrical spin frustration has been recently recognized to be one of the keys  for 
the ME multiferroic materials. 
The frustrated system, owing to the vast degeneracy arising from competing magnetic 
interactions,  often displays complex magnetic orders at low temperatures like
non-centrosymmetric noncollinear or long-period-modulated collinear magnetic orders. 
These exotic magnetic structures sometimes break the crystal symmetry and can be an origin of the ferroelectricity.

CuFeO$_2$ (CFO), which is one of the model materials of a triangular lattice 
antiferromagnet, has been extensively investigated as a geometrically frustrated spin 
system~\cite{Mitsuda1991nds,Petrenko2005rmp}.
Since the recent discovery of ferroelectricity in the first
magnetic-field-induced phase~\cite{Kimura2006ibi}, CFO has attracted more attention as a distinct class
 of ME multiferroics.
Three possible mechanisms~\cite{Jia2007mto} have been proposed for the coupling between magnetic and electric
 moments for the ME multiferroic materials:
(1) magnetostriction, (2) spin current~\cite{Katsura2005sca}, and (3) spin-dependent $d$-$p$ hybridization.
CFO is the first case that the electric polarization is not explained by either
the first or the second model but is possibly explained by 
 the third model, spin-dependent hybridization between transition metal ion Fe $3d$ and ligand ion 
 O $2p$~\cite{Arima2007fib,Nakajima2008iom}.

 Here we show direct evidence for a long range order of the $d$-$p$ hybridization 
 using  soft x-ray resonant Bragg diffraction.
Superlattice reflection  0, 1-2$q$, 0 ($2q=0.828$ is twice of the magnetic wave number along the $b$ axis) 
has been found at Fe L$_{2,3}$ absorption edges.
The integrated intensity of this reflection was measured as a function of temperature,  
x-ray polarization,  and x-ray energy.   
Our experimental findings definitely show that the $3d$ orbital state of Fe ions has a long range order in the ferroelectric state.

CFO has the delafossite structure space group $R\overline{3}m$ at room temperature.
The structure consists of Fe$^{3+}$ hexagonal layers along the $c$ axis, which are separated 
by intervening two layers of oxygen and one layer of Cu$^{1+}$ 
as shown in Figure 1 (a).
CFO undergoes two successive magnetic transitions at low temperatures 
in the absence of external magnetic field.
At $T_{N1}=14$ K the paramagnetic phase turns to be the sinusoidally modulating 
partially disordered (PD) state with a wave vector 
$Q=(0,q,\frac{1}{2})_m$, $q \sim \frac{2}{5}$ 
and then at $T_{N2}=11$ K, it turns to be the collinear four-sublattice (4SL) ground state 
$\uparrow\uparrow\downarrow\downarrow$ with a wave vector 
$Q=(0, \frac{1}{2},\frac{1}{2})_m$~\cite{Mitsuda1998pdp}.  
Here the suffix $m$ means that these 
indices are based on the monoclinic space group $C2/m$ which represents 
the crystal structure below the temperature $T<T_{N1}$~\cite{Ye2006ssc}.
Hereafter the monoclinic notation is used unless specified.

A spontaneous electric polarization along the direction parallel to the $b$ axis appears  
in the first field-induced phase by applying a magnetic field along the hexagonal 
$c$ axis~\cite{Kimura2006ibi} or substituting a few percentage of Fe ions
with non-magnetic (Al or Ga) ions~\cite{Seki2007ifi, Terada2008gsf} in the absence of external magnetic field.
Nakajima et al.\ have clarified that the ferroelectric phase
has a proper helical magnetic order with an incommensurate periodicity~\cite{Nakajima2007sni} and 
that the magnetic chirality, \textit{left-handed} or \textit{right-handed},  directly corresponds to 
the polarity ($+$ or $-$) of the spontaneous electric polarization using
polarized neutron diffraction~\cite{Nakajima2008epi}.
Figure 1 (a) shows the left-handed proper helical magnetic order in 
the ferroelectric and incommensurate magnetic (FE-ICM) ordered  phase, where the electric polarization
vector $\mathbf{P}$ points to the negative direction along the $b$ axis.
Note that the space group in the FE-ICM phase is not $C2/m$ anymore because the magnetic order breaks 
the $C$-face-centred structure.
Moreover the mirror symmetry perpendicular to the $b$-axis is to be broken, too.
This symmetry breaking is essential for realizing the spontaneous electric polarization.
Figure 1 (b) shows the impurity--temperature magnetic phase 
diagram~\cite{Terada2009mpd} of CuFe$_{1-y}$Ga$_y$O$_2$ 
consisting of four magnetic phases: 4SL, PD, FE-ICM, and oblique PD (OPD) phases 
and the paramagnetic (PM) phase.  

\begin{figure}[htp]
\includegraphics[width=75mm]{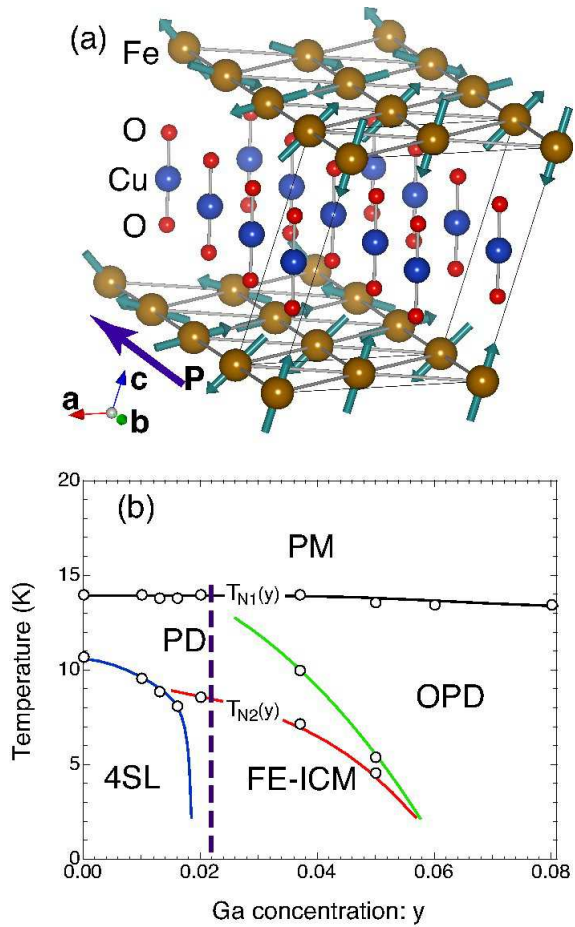}
\caption{\label{fig:1} 
(Color online) Appearance of ferroelectricity in CuFeO$_2$.
(a) Atomic configuration and magnetic structure of CuFeO$_2$.
Below temperature $T<T_{N1}$, the crystal structure is the monoclinic space group $C2/m$.
The unit cell is drawn with a setting of unique axis $b$ and cell choice 1 
of the monoclinic space group $C2/m$.  Black lines show the unit cell.
The \textit{left-handed} proper helical magnetic structure in ferroelectric phase is illustrated by green arrows
at iron ions.  When the magnetic chirality is left-handed, the electric polarization vector points the
negative direction along the $b$ axis.
(b) Magnetic phase diagram of CuFe$_{1-y}$Ga$_y$O$_2$ drawn as functions of temperature and the content of Ga ions.
Data are taken from Ref.~\onlinecite{Terada2009mpd}.  The sample used for the soft x-ray diffraction 
has 2.2 \% Ga, indicated by a dash line. 
}
\end{figure}

 There have been presented mainly three possible mechanisms to explain the ferroelectricity in 
 the ME multiferroic materials.
 Among them, the magnetostriction model induces the electric polarization
 only for a commensurate magnetic order but not for an incommensurate magnetic order.
 The spin-current model~\cite{Katsura2005sca} conflicts with
 the above experimental findings because it does not predict that the electric polarization vector is
 parallel to the magnetic wave vector.   
 Most plausible mechanism for CFO is the spin-dependent \textit{d}-\textit{p} hybridization model~\cite{Arima2007fib}.
 According the model, the spin-orbit coupling in Fe ions gives rise to a spatial variation of metal-ligand (Fe-O) 
 hybridization associated with the proper helical magnetic structure.
 It predicts  (1) that a charge transfer between Fe and O ions spatially varies together 
 with a variation of the  hybridization, 
 (2) that an electric polarization is induced parallel to the magnetic wave vector (along the $b$ axis),
(3) that the polarity of the electric polarization corresponds to the magnetic chirality, 
and (4) that the electric polarization has 
 spatial modulations of wave numbers of $2q$ and $4q$.
 Here $q$ is the wave number of the proper helical magnetic structure. 
 Actually, Nakajima et al.\ have found superlattice reflections having $2q$ modulation in the FE-ICM phase of
 CuFe$_{1-y}$Al$_y$O$_2$ ($y=0.0155$) using the non-resonant x-ray diffraction~\cite{Nakajima2008iom}. 
They have pointed out that the Fourier-transform spectrum of the spatial modulation based 
 on the model is consistent
 with the experimental results including the intensity ratio of the superlattice reflections.
 On the other hand, the conventional magnetostriction mechanism 
 does not fit the experimental results.
 This result supports the spin-dependent \textit{d}-\textit{p} hybridization model 
 in the FE-ICM phase of CFO but does not provide any direct evidence
   of a spatial variation of the metal-ligand hybridization.
 
   We have performed soft x-ray resonant diffraction to elucidate the mechanism of
 the ferroelectricity in CFO at Fe $L_{2,3}$ absorption edges.
 Conventional x-ray diffraction is commonly used to determine structures of a variety of 
 materials ranging from inorganic compounds to biomolecules.  
 On the other hand, resonant x-ray diffraction (RXD) provides a powerful tool 
 to investigate a variety of ordered states, such as magnetic, charge or orbital orders.
  Most important feature of RXD is the strong dependence on the \textit{polarization} state of photons.
This property was first demonstrated by Templeton and Templeton in 1982~\cite{Templeton1982xda}.
RXD occurs as the second perturbation in the interaction between photons and atomic
electrons, where photons exchange the energy with the electron system by electric or magnetic multipole
transitions.
Passing through the intermediate states (unoccupied states), the x-ray scattering is quite sensitive 
to the local electric environment of the resonant element.
The scattering cross-section, then, is described by 
the atomic tensors~\cite{Lovesey2005epo} or 
the anisotropy of x-ray susceptibility~\cite{Dmitrienko2005pao}.
 
 Single crystals of CuFe$_{1-y}$Ga$_y$O$_2$ with $y=0.022$ were grown 
by the floating-zone technique.  They were cut into disks with 2 mm in thickness
and polished to remove the surface roughness.
The surface is perpendicular to the monoclinic $b$ axis.
We have performed RXD experiments for $L_{2,3}$ absorption edges of Fe using a soft x-ray diffractometer built
 at beam line 17SU~\cite{Ohashi2007poh} in SPring-8. 
 The sample was mounted  in a liquid $^4$He flow-type cryostat.
  The incident x-ray energy was tuned  by the grating apparatus. 
  The polarization of the incident photons was switched 
  by the electromagnet of the undulator~\cite{Shirasawa2004ooa} and that of the diffracted photons
  was analyzed by a W/B$_4$C multilayer.  The bilayer thickness of the multilayer is adjusted to the 
  wavelength for Fe $L_3$ edge at Bragg angle $2\theta=90$ degrees.
Note that  the degree of linear polarization of the x-ray beam is determined by
the components perpendicular ($\sigma$) and parallel ($\pi$) to the scattering plane, 
 which is spanned by the incident and diffracted propagation vectors  ${\mathbf k}_i$ and ${\mathbf k}_f$.
 Our insertion device for the incident beam can control the polarization of the incident beam, but not perfectly. 
 Thus, the actual linear polarization in the experiments 
 has 98.8\%  for $\pi$ incident beam  and 83\%  for $\sigma$ incident beam, respectively.
Data shown in Figure 2 are corrected with these factors.

\begin{figure}
\includegraphics[width=85mm]{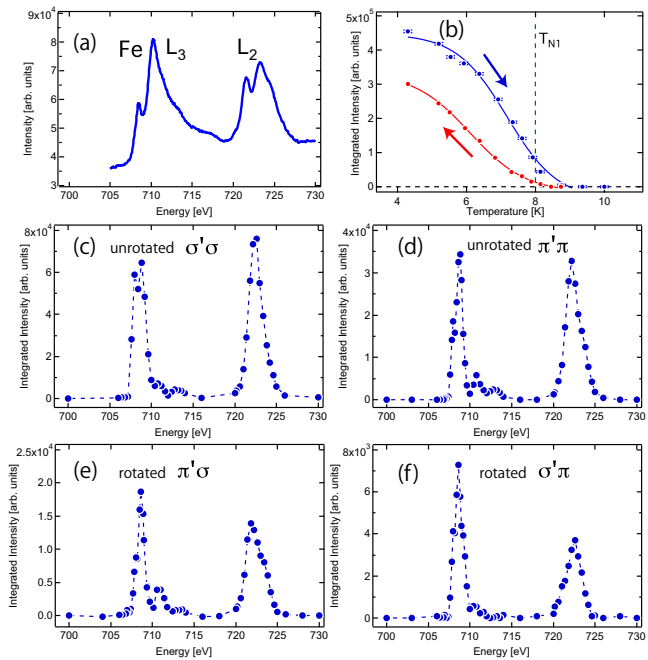}
\caption{\label{fig:2} 
(Color online) Energy spectra near the Fe $L_{2,3}$ absorption edges.
All data were measured at temperature $T=4.2$ K except panel b. 
(a) XAS near the Fe $L_3$ and $L_2$ absorption edges was obtained in the total x-ray fluorescence yield mode.
(b) The integrated intensity  of superlattice reflection 0, $1-2q$, 0 was measured
as a function of temperature.  The energy of x-ray beam was E = 708.6 eV\@.
The measurement was carried out in two ways, first increasing temperatures (a blue line) from 4 K up to 20 K, and then 
decreasing temperatures (a red line).
Integrated intensities of superlattice reflection 0, $1-2q$, 0 for unrotated channels 
 (c) $\sigma'\sigma$  and  (d) $\pi'\pi$
and for rotated channels (e) $\pi'\sigma$  and (f) $\sigma'\pi$ are shown. 
Azimuthal angle was fixed for these measurements as the hexagonal $c$ axis was about 20 degrees
off from normal vector of the scattering plane.
}
\end{figure}

The results of RXD measured at Fe $L_{2,3}$ absorption edges are summarized 
in Figure 2.
Panel \textbf{a} shows the x-ray absorption spectrum (XAS) of Fe $L_{2,3}$ edges obtained 
in the total x-ray fluorescence yield mode.
The maxima of Fe absorption appear at 710.3 eV and 723.3 eV, corresponding to 
$L_3$ and $L_2$ absorption
edges, respectively.
Double-peak structures observed both at $L_{3}$  and $L_{2}$ edges 
originate from the crystal-field splitting of Fe ions~\cite{Crocombette1995xsa}.
Peaks observed at lower and higher energies correspond to the transition to the
$3d$ $t_{2g}$ and $e_{g}$ states, respectively, for both $L_{3}$  and $L_{2}$ edges.
Panel \textbf{b} shows the integrated intensity observed at $E=708.6$ eV as a function of temperature. 
It gradually decreases with increasing temperatures and disappears around 9 K 
and it gradually increases with decreasing temperatures after the sample is warmed up to 20 K, where the system is 
in the paramagnetic phase. 
As shown in Figure 1 b, the sample ($y=0.022$) 
has the magnetic transition at $T_{N2}=8$ K\@.
The intensity appears differently in two ways 
because the magnetic transition at 
$T_{N2}$ is the first order and the monoclinic domains appear differently for each time in cooling process 
from the PM phase.

Energy spectra of the integrated intensity of superlattice reflection 0, $1-2q$, 0
measured at  temperature $T=4.2$ K 
for polarization-unrotated channels  $\sigma'\sigma$ and  $\pi'\pi$ and 
rotated channels  $\pi'\sigma$ and  $\sigma'\pi$ are displayed in Figure 2, panels \textbf{c} to \textbf{f}.
Here $\sigma$ ($\sigma'$) and $\pi$ ($\pi'$) show the polarization in the scattering for incident
(diffracted) photons.
We have clearly found superlattice reflection 0, $1-2q$, 0
in the FE-ICM phase for the x-ray polarization channels (panels c, d, e and f) at both $L_{3}$  and $L_{2}$ edges.
Note that the maximum of the peak at $L_{3}$ edge for each panel corresponds to the resonant transition 
to the $t_{2g\downarrow}$ orbit according to the lower-energy peak of the XAS\@.
Most important feature found in these energy spectra is that 
superlattice reflection 0, $1-2q$, 0
is observed for polarization-rotated
channels $\pi'\sigma$ and $\sigma'\pi$.  
The rotation of  polarization is forbidden for conventional x-ray diffraction.
It is allowed only when the local electronic state of the resonant ion, which
is expressed by the second or higher rank of tensors, is aspherical and has a 
periodic spatial modulation as well.
Thus the rotation of the polarization found here
clearly proves that the Fe $3d$ orbital state has a motif of
a wave vector $2q$ along the $b$ axis.

Let us discuss what picture we deduce for the appearance of ferroelectricity from our findings.
The proper helical magnetic structure found in the polarized neutron diffraction 
 conflicts with the crystal symmetry and breaks it.
Most probable mechanism of this symmetry-breaking is the spin-dependent hybridization 
between the Fe $3d$ state and the O $2p$ state as described in Ref~\onlinecite{Arima2007fib}. 
The spin-orbit interaction in Fe ions lifts the degeneracy
of the $t_{2g\downarrow}$ orbits of the Fe high spin state $^6S$ 
as a perturbation and 
this lifting causes the spatial periodic modulation of the Fe $3d$ state along the $b$ axis, 
coupling with the magnetic structure.
Our findings provide  direct evidence of the modulation of the $t_{2g\downarrow}$ orbit
 in the FE-ICM phase having wave number $2q$.
 Arima~\cite{Arima2007fib} has proposed that
 a charge transfer between the Fe and the O ions appears according to the hybridization.
 However, we have  no evidence for  a 
 charge-transfer modulation so far~\cite{Terada2010eaw}.
If it exits,  some resonant signature should 
have been observed at Fe $K$ edge.
For example we have found that forbidden reflection 010  
in the 4SL phase of pure CuFeO$_2$ is strongly enhanced 
at Fe $K$ edge due to the charge disproportionation~\cite{Terada2010cda}.
Thus our findings agree with the spin-dependent \textit{d}-\textit{p} hybridization model
but do not support a charge-transfer modulation.

Hereby we would rather propose that the position of O ions varies with 
the hybridization between Fe $3d$ and O $2p$ via spin-orbit coupling
than a charge-transfer modulation.
This spatial modulation breaks the crystal symmetry and invokes 
the ferroelectricity in this compound.
Actually, we have observed that superlattice reflection 0, $1+2q$, 0 
appears in non-resonant energies around Fe $K$ absorption edge~\cite{Terada2010eaw}. 
On the other hand, we have not found any signature of resonant intensity for reflection 
0, $1-2q$, 0 at O $K$ absorption edge, yet.
Polishing the sample may damage the surface structure associated with an oxygen loss. 
Such a bad surface condition might prevent the observation.

 In summary we have performed resonant soft x-ray diffraction on the 
 multiferroic material
 CuFe$_{1-y}$Ga$_y$O$_2$, $y=0.022$ and found that
 a superlattice reflection  0, 1-2$q$, 0 at Fe L$_{2,3}$ absorption edges appears in the FE-ICM phase
 and that the x-ray polarization for  this reflection rotates.
 These findings provide
 direct evidence for the spatial periodic modulation of the Fe 3\textit{d} orbital state.

\begin{acknowledgments}
We thank M. Taguchi for valuable discussions.
 This work was partly  supported by Grants-in-Aid for Scientific Research  
(A) No. 21244049 from JSPS and for Young Scientists (B) No. 2074209 from JSPS.
The synchrotron radiation experiments were performed at beam line 17SU in SPring-8 with the approval of RIKEN 
(Proposal No. 20070129, 20080045, and 20090012).
\end{acknowledgments}

\nocite{*}

\bibliography{cufeo2}

\end{document}